\documentclass{svjour2}                    
\smartqed  
\usepackage{graphicx}

\def\PRA{{Phys. Rev.} A }
\journalname{}
\begin{document}

\title{On the discrimination between classical and quantum states
\thanks{This work has been supported by SanPaolo Foundation, Quantum Candela EU Project}
}


\author{G. Brida \and  M. Bondani\\
        I. P. Degiovanni\and M. Genovese\\
        M. G. A. Paris \and I. Ruo Berchera\\
        V. Schettini}


\institute{G. Brida \at
              Istituto Nazionale di Ricerca Metrologica, I-10135,
Torino, Italia \\
              \email{g.brida@inrim.it}
           \and
           M. Bondani \at
              National Laboratory for Ultrafast and Ultraintense
              Optical Science,\\ C.N.R.-I.N.F.M., I-22100 Como, Italia\\
              C.N.I.S.M., U.d.R. Como, I-22100 Como, Italia\\
              \email{maria.bondani@uninsubria.it}
           \and
           I. P. Degiovanni \at
              Istituto Nazionale di Ricerca Metrologica, I-10135,
Torino, Italia \\
              \email{i.degiovanni@inrim.it}
           \and
          M.Genovese \at
              Istituto Nazionale di Ricerca Metrologica, I-10135,
Torino, Italia \\
              \email{m.genovese@inrim.it}   \and
              M. G. A. Paris \at
              Dipartimento di Fisica, Universit\`a degli Studi di Milano, I-20133 Milano, Italia\\
              C.N.I.S.M., U.d.R. Milano Universit\`a, I-20133 Milano, Italia\\
              I.S.I. Foundation, I-10133 Torino, Italia \\ \email{matteo.paris@fisica.unimi.it}
           \and
         I. Ruo Berchera \at
              Istituto Nazionale di Ricerca Metrologica, I-10135,
Torino, Italia \\
              \email{i.ruoberchera@inrim.it} \and
              V. Schettini \at
              Istituto Nazionale di Ricerca Metrologica, I-10135,
Torino, Italia \\
\email{v.schettini@inrim.it}}


\maketitle

\begin{abstract}
With the purpose to introduce a useful tool for researches concerning
foundations of quantum mechanics and applications to quantum
technologies, here we study three quantumness quantifiers for bipartite
optical systems: one based on sub-shot-noise correlations, one related
to antibunching and one springing from entanglement determination. The
specific cases of parametric down conversion
seeded by thermal, coherent and squeezed states are discussed in detail.
\end{abstract}

\section{Introduction}
\label{intro} The discrimination between quantum and classical
states  \cite{vax,gen05,A,pg,san}, beyond its very important and
deep conceptual relevance, has also recently received much attention
due to the development of quantum technologies.
On the one hand, it represents a fundamental point for the
studies concerning the transition between quantum and classical
world, one of the most intriguing research sectors in the
foundations of physics; on the other hand it is a tool of the utmost
importance when comparing results that can be achieved with quantum and
classical protocols.

These studies have concerned various physical systems \cite{vax}
ranging from quantum optical states \cite{pg} to mesons \cite{gen1}
or solid state devices \cite{ss} and, recently, have pointed to the
need of an operational approach linked to measurements schemes
\cite{si04,si05,sa05,si03,R5,R6,R7,R8,R9,R10,R11,R12,R13,R14,R15,al1,al2,ch}.
Considering the experimental interest in the frame of quantum optics
\cite{ch,dem,gr,b}, in a recent paper \cite{nos1} we have considered
three different ``quantumness" quantifiers applied to the study of
quantum-classical transition in a thermal seeded parametric down
conversion (PDC): a work that, a part its specific application, has
a more general interest since it presents an idea that can find
generalization to various physical systems.

Here we want to extend this first study by considering, as a further
example, the application of these three quantifiers to PDC seeded by
coherent and squeezed vacuum states, comparing them with what
we achieved for thermal seeded case. This allows a more general
understanding of the hierarchy of these three quantifiers that
emerged in \cite{nos1} and, due to the easy realizability of these
states, paves the way toward a general experimental test of these
theoretical results.

The structure of the paper is the following. After an introduction to
seeded PDC (section 2), we consider the sub-shot noise measurement
(section 3), Lee's criterion (section 4) and entanglement (section
5). Finally, a general discussion of the results is presented in
section 6.
\section{Introduction to seeded parametric down-conversion }
\label{s:MMTPDC}
\par
The evolution of a quantum system induced by the interaction
Hamiltonian describing the PDC process for a single pair of coupled
modes is described by the unitary operator $U=\exp(i \kappa
{a_A}{a_B} + h.c.)$, where $\kappa=|\kappa| e^{i \varphi}$ is the
coupling constant and $a_A$ and $a_B$ are the annihilation operators
for photons on modes A and B, respectively. We consider the PDC
process seeded by two single mode input states
${\rho}_{in}={\rho}_{A}\otimes {\rho}_{B}$. In particular as seeds
fields on both A- and B- modes we consider the three simplest
Gaussian states, namely thermal states, coherent states, and vacuum
squeezed states. For the thermal case (T) the input state of the
single channel is a mixed incoherent superposition.
\begin{equation}\label{rhoin(T)}
{\rho}^{(T)}_{j}=\sum_{n=0}^{\infty}P_{j}(n)~ |n\rangle_{j}\langle n|\:,\\
\end{equation}
where $j=A,B$ and $\left| n \right\rangle_{j}$ denotes the Fock
number basis for the single mode of the $j$-arm, the thermal
probability distribution of the input being
$P_{j}(n)=\mu_{j}^{n}(1+\mu_{j})^{-n-1} ,$ where $\mu_{j}$ is the
average photon number.

In the case of coherent seeding (C) the state in the single channel
is obtained by the action of the displacement operator
$D(\alpha_{j})=e^{i( \alpha_{j}a_{j}+
\alpha^{*}_{j}a_{j}^{\dagger})}$ on the vacuum state
\begin{equation}\label{rhoin(C)}
{\rho}^{(C)}_{j}=D(\alpha_{j})|0\rangle_{j}\langle0|D^{\dagger}(\alpha_{j})
\end{equation}
with $\alpha_{j}= \sqrt{M_{j}}\:e^{i\gamma_{j}}$ is the complex
displacement parameters and $M_{j}$ represents the mean number of
photon of the state.

For the squeezed vacuum seeds (S) the input state of the single
channel is given by
\begin{equation}\label{rhoin(S)}
{\rho}^{(S)}_{j}=S(\xi_{j})|0\rangle_{j}\langle0|S^{\dagger}(\xi_{j})
\end{equation}
where  the squeezing operator is $S(\xi_{j})=e^{i(
\xi_{j}a^{2}_{j}+\xi_{j}^{*} a^{\dagger 2}_{j})}$, $\xi_{j}=
|\xi_{j}|\:e^{i\zeta_{j}}$ being complex parameters.

The density matrix at the output of the PDC interaction is given by
\begin{equation}\label{rhoout}
{\rho}_{out} = {U}{\rho}_{in}{U}^{\dagger}.
\end{equation}

Conversely in the interaction picture the output field modes are
given by ${A}_{j}={U}^{\dag}{a}_{j}{U}$, {\em i.e}
\begin{equation}
{A}_{j}=\sqrt{N+1}\: {a}_{j}+ e^{i \varphi}\sqrt{N}\:
{a}_{j'}^{\dag} ~~ (j,j'=A,B , ~j \neq j') \label{aout}
\end{equation}
where $N= \sinh^2 |\kappa |$ is the mean number of photons of the
PDC spontaneous emission.

The first moments of the photon number distribution in the case of
thermal seeds are
 \cite{DBA07}
\begin{eqnarray} \label{averages(T)}
\langle {n}^{(T)}_{A} \rangle &=&\mu_{A}+ N(1+ \mu_{A}+ \mu_{B})
\nonumber\\
\langle {n}^{(T)}_{B} \rangle &=& \mu_{B}+ N(1+ \mu_{A}+ \mu_{B})
\end{eqnarray}
where, $n_{j}= a_{j}^{\dag} a_{j} $, $\langle {O} \rangle=
\mathrm{Tr}[{O}{\rho}_{out}]$.

When the process is seeded by coherent fields we have:
\begin{eqnarray} \label{averages(C)}
\langle {n}^{(C)}_{A} \rangle &=&M_{A}+ N(1+ M_{A}+
M_{B})+2\sqrt{N(N+1)}\sqrt{M_{A}M_{B}}
\cos(\gamma_{A}+\gamma_{B}-\varphi)
\nonumber\\
\langle {n}^{(C)}_{B} \rangle &=&M_{B}+ N(1+ M_{A}+
M_{B})+2\sqrt{N(N+1)}\sqrt{M_{A}M_{B}}
\cos(\gamma_{A}+\gamma_{B}-\varphi)
\nonumber\\
\end{eqnarray}
Contrary to the thermal case, here the intensity of the fields is
partially modulated by the phase value of the seeding fields with
respect to the phase induced by the PDC process when both the seeds
are nonzero (i.e. $M_{A},M_{B}\neq 0$). Eventually when vacuum
squeezed input beams are considered, we obtain
\begin{eqnarray} \label{averages(S)}
\langle {n}^{(S)}_{A} \rangle &=&N_{A}+ N(1+ N_{A}+
N_{B})\nonumber\\
\langle {n}^{(S)}_{B} \rangle &=&N_{B}+ N(1+ N_{A}+ N_{B})\:,
\nonumber\\
\end{eqnarray}
where $N_{j}=\sinh^2|\xi_{j}|$ is the average number of photons of
the input state in the single channel $j$, whereas, surprisingly,
the phases of the squeezing operators do not play any role in the
photon number. Notice that the case of vacuum inputs, ${\rho}_{in}=
|0\rangle\langle 0|_{A}\otimes|0\rangle\langle 0|_{B}$, corresponds
to spontaneous PDC, {\em i.e.} to the generation of twin-beam.
\section{Sub-Shot Noise Measurement}

The shot-noise limit (SNL) in any photodetection process is defined
as the lowest level of noise that can be obtained by using
semiclassical states of light that is, Glauber coherent states. If
one measures the photon numbers in two beams and evaluates their
difference, the SNL is the lower limit of noise that can be reached
when the beams are classically correlated. On the other hand, when
intensity correlations below the SNL are observed, we have a genuine
nonclassical effect. We consider a simple measurement scheme where
$A$ and $B$ single mode beams at the output of the PDC interaction
are individually measured by direct detection (considering in this
section ideal detectors with unitary efficiency \cite{ruo}). The
resulting $A$ and $B$ photon counts, which are correlated, are
subtracted from each other to demonstrate quantum noise reduction in
the difference of photon counts. In order to observe a violation of
the SNL we must have
\begin{equation} \label{ssnA}
 \langle [\Delta
(n_{A}-n_{B})]^{2}\rangle <  \langle {n}_{A} \rangle + \langle
{n}_{B} \rangle  \:
\end{equation}
where $\langle [\Delta (n_{A}-n_{B})]^{2}\rangle$ is the variance of
this difference, and $\langle {n}_{A} \rangle + \langle {n}_{B}
\rangle$ is the SNL, \textit{i.e.} the quantity that would be
obtained for uncorrelated coherent beams.

In particular, the condition in Eq. (\ref{ssnA}) reduces to
\begin{eqnarray} \label{ssn(T)}
N>\frac{(\mu_A^2 + \mu_B^2)}{2  (1+ \mu_A + \mu_B)}
\end{eqnarray}
for the thermal seeds.

In the case of coherent seeds if the phases satisfy
$\cos(\gamma_{A}+\gamma_{B}-\varphi)\geq 0$ the condition in Eq.
(\ref{ssnA}) is always fulfilled irrespective of the value of $N$,
$M_{A}$, $M_{B}$, while if $\cos(\gamma_{A}+\gamma_{B}-\varphi)<0$
the condition is fulfilled only when
\begin{equation} \label{ssn(C)}
  N>\frac{4
M_{A}M_{B}\cos^2(\gamma_{A}+\gamma_{B}-\varphi)}{1+2(M_{A}+M_{B})+(M_{A}+M_{B})^2-4M_{A}M_{B}\cos^2(\gamma_{A}+\gamma_{B}-\varphi)}.
\end{equation}

In the case of squeezed vacuum seeds the condition is
\begin{eqnarray} \label{ssn(S)}
N>\frac{N_{A}(1+2N_{A})+ N_{B}(1+2N_{B})}{2  (1+ N_{A}+N_{B})}.
\end{eqnarray}
It is interesting to notice that, in the case of thermal and squeezed
input state, there always exists a threshold between the sub-shot-noise
and the classical regime, which can be explored by controlling the
intensities of the seeds. The behavior of the coherent case is
different because, upon properly adjusting the phases, the sub-shot noise
condition holds whatever the intensities of the seeds. In an
experiment in which the phases of seeds $\gamma_{A},\gamma_{B}$ and
that of PDC process $\varphi$ are not locked, one expects, on
average, a null value of the cosine in Eq. (\ref{ssn(C)}) and
therefore a permanent SSN condition. It is helpful to define a
parameter, $\mathcal{P}_{SSN}$ quantifying the amount of violation
of the SNL
\begin{equation} \label{Pssn}
\mathcal{P}_{SSN}= 1- \frac{\langle [\Delta
(n_{A}-n_{B})]^{2}\rangle}{\langle {n}_{A} \rangle + \langle {n}_{B}
\rangle}.
\end{equation}
$\mathcal{P}_{SSN}= 0$ corresponds to the SNL, and the sub-shot
noise condition corresponds to $0 < \textit{P}_{SSN}  \leq 1$. For
the state $\rho_{out}$, in the case of thermal seeds we have

\begin{equation} \label{Pssn(T)}
\mathcal{P}^{(T)}_{SSN}= \frac{2 ~ \mu_{PDC} (1+ \mu_A + \mu_B)-
\mu_A^2 - \mu_B^2}{2 ~ \mu_{PDC} (1+ \mu_A + \mu_B)+ \mu_A + \mu_B},
\end{equation}
thus the maximal violation of SNL is achieved by the twin-beam
($\mu_A=\mu_B=0$), and by increasing the magnitude of, at least, one
of the seeding field the SNL is reached.

For coherent input beams the amount of violation is
\begin{equation} \label{Pssn(C)}
\mathcal{P}^{(C)}_{SSN}= \frac{ 2N(1+ M_{A} + M_{B})+
4\sqrt{N(N+1)}\sqrt{M_{A}M_{B}}\cos(\gamma_{A}+\gamma_{B}-\varphi)}{2N(1+
M_{A} + M_{B})+
4\sqrt{N(N+1)}\sqrt{M_{A}M_{B}}\cos(\gamma_{A}+\gamma_{B}-\varphi)+M_{A}
+ M_{B}},
\end{equation}
and also in this case the limit value of 1 is reached again by the
twin-beams in the spontaneous emission ($M_{A}=M_{B}=0$). The SNL
threshold $\mathcal{P}^{(C)}_{SSN}=0$ is obtained when the numerator
of Eq. (\ref{Pssn(C)}) is zero, leading to the solutions presented
in Eq.(\ref{ssn(C)}).

Finally, for the squeezed beams the parameter is
\begin{equation} \label{Pssn(S)}
\mathcal{P}^{(S)}_{SSN}= \frac{ 2N(1+ N_{A} + N_{B})-
2N_{A}(1+N_{A})-2N_{B}(1+N_{B})}{2N(1+ N_{A} + N_{B})+ N_{A} +
N_{B}}.
\end{equation}

We notice that for all the cases, it can be shown that
$\langle[\Delta (n_{A}-n_{B})]^{2}\rangle$ is equal to the sum of
the mean fluctuation of the two input seeding states. Therefore,
$\mathcal{P}_{SSN}$ always assumes the form
\begin{equation} \label{}
\mathcal{P}_{SSN}\equiv1- \frac{\langle [\Delta
(n_{A}-n_{B})]^{2}\rangle}{\langle {n}_{A} \rangle + \langle {n}_{B}
\rangle}=\frac{\langle {n}_{A} \rangle + \langle
{n}_{B}\rangle-\langle[\Delta
n_{A}]^2\rangle_{\rho_{in}}-\langle[\Delta
n_{B}]^2\rangle_{\rho_{in}}}{\langle {n}_{A} \rangle + \langle
{n}_{B} \rangle},
\end{equation}
where, with obvious notation, $\langle {O}
\rangle_{\rho_{in}}= \mathrm{Tr}[{O}{\rho}_{in}]$.

\section{Lee's Criterion}
Another interesting criterion of nonclassicality was derived by Lee
\cite{Lee1,Lee2}, and it is the two-mode generalization of the well
known nonclassicality criterion for single mode beam $ \langle n
(n-1) \rangle - \langle n \rangle^2 < 0 $ \cite{Man79}. The Lee's
criterion states that a bipartite system is nonclassical if the
inequality
\begin{equation} \label{leeA}
\langle n_{A}(n_{A}-1)\rangle + \langle n_{B}(n_{B}-1)\rangle - 2
\langle n_{A} n_{B}\rangle < 0
\end{equation}
is satisfied. It is noteworthy to observe that the "Lee's
nonclassicality" corresponding to Eq. (\ref{leeA}), implies the
negativity of the Glauber-Sudarshan P-function \cite{Lee1,Lee2}.
Once we consider the state $\rho_{out}$, the condition in Eq.
(\ref{leeA}) for seeded PDC is achieved when
\begin{equation} \label{lee(T)}
N>\frac{\mu_A^2 + \mu_B^2 - \mu_A  \mu_B}{(1+ \mu_A + \mu_B)} ,
\end{equation}
while in the case of coherent seeds, we have that, when the phases
satisfy $\cos(\gamma_{A}+\gamma_{B}-\varphi) \geq 0$, $N>N_{-}$,
while when $\cos(\gamma_{A}+\gamma_{B}-\varphi)<0$, $N>N_{+}$ with
\begin{equation} \label{}
N_{\pm}=\frac{
4g\cos^{2}r+ab\pm2[g\cos^{2}r(4g\cos^{2}r+2ab+a^2)]^{1/2}}{2[b^2-4g\cos^{2}r]}
\end{equation}
and $g=M_{A}M_{B},$ $a=(M_{A}-M_{B})^{2},$ $b=1+M_{A}+M_{B},$
$r=\gamma_{A}+\gamma_{B}-\varphi$.

For the squeezed thermal seeds we obtain
\begin{equation} \label{lee(S)}
N>\frac{N_{A}+N_{B}+3N^2_{A}+3N^2_{B}-2N_{A}N_{B}}{2(1+N_{A}+N_{B})}.
\end{equation}
It is noteworthy to observe that the Lee condition is stricter than
the sub-SNL: always exists a threshold between classicality and
nonclassicality for the Lee's criterion.

Analogously to the sub-SNL case we define a parameter quantifying
the amount of violation of the classicality bound as
\begin{equation} \label{Plee}
\mathcal{P}_{Lee}= 1- \frac{\langle [\Delta
(n_{A}-n_{B})]^{2}\rangle+ (\langle {n}_{A} \rangle - \langle
{n}_{B} \rangle)^2}{\langle {n}_{A} \rangle + \langle {n}_{B}
\rangle}.
\end{equation}
$\mathcal{P}_{SSN}= 0$ corresponds to the bound of the Lee
nonclassicality region, and the Lee nonclassicality condition
corresponds to $0 < \mathcal{P}_{Lee} \leq 1$. For the thermal seeds
we obtain
\begin{equation} \label{Plee(T)}
\mathcal{P}^{(T)}_{Lee}=2 ~\frac{ N (1+ \mu_A + \mu_B)- \mu_A^2 -
\mu_B^2+\mu_A \mu_B}{2 ~ N (1+ \mu_A + \mu_B)+ \mu_A + \mu_B}.
\end{equation}
Thus, the maximal violation of Lee's criterion
($\mathcal{P}_{Lee}=1$) is achieved by the twin-beam
($\mu_A=\mu_B=0$), and by increasing the magnitude of, at least, one
of the seeding field the classicality bound is reached.

For coherent input beams, the Lee parameter is
\begin{equation} \label{Plee(C)}
\mathcal{P}^{(C)}_{Lee}=\frac{ 2N(1+ M_{A} + M_{B})+
4\sqrt{N(N+1)}\sqrt{M_{A}M_{B}}\cos(\gamma_{A}+\gamma_{B}-\varphi)-(M_{A}
- M_{B})^{2}}{2N(1+ M_{A} + M_{B})+
4\sqrt{N(N+1)}\sqrt{M_{A}M_{B}}\cos(\gamma_{A}+\gamma_{B}-\varphi)+M_{A}
+ M_{B}}.
\end{equation}
It shows a maximum non-classical violation of Lee criterion
($\mathcal{P}^{(C)}_{Lee}=1$) when the intensities of seeds are
null. Eventually, in the case of squeezed seeding one obtains
\begin{equation} \label{Plee(S)}
\mathcal{P}^{(S)}_{Lee}=\frac{ 2N(1+ N_{A} + N_{B})-
2(N^{2}_{A}+N^{2}_{B})-(N_{A}-N_{B})^{2}-N_{A}-N_{B}}{2N(1+ N_{A} +
N_{B})+ N_{A} + N_{B}}.
\end{equation}

\section{Entanglement}

The downconversion process is known to provide pairwise entanglement
between $A-$ and $B-$ beams. In the spontaneous process, as well as
in the case of coherent and vacuum squeezed seeds the output state
is entangled for any value of $N \neq 0$ whereas in the case of a
thermally seeded PDC the degree of entanglement crucially depends on
the intensity of the seeds, as shown in \cite{DBA07}. In fact, as
$\rho_{out}$ is a Gaussian state (since thermal states are Gaussian
and the PDC Hamiltonian is bilinear in the field modes) its
entanglement properties can be evaluated by checking the positivity
of the partial transpose (PPT condition), which represents a
sufficient and necessary condition for separability for Gaussian
pairwise mode entanglement \cite{simon00}.

In order to check whether and when the state ${\rho}_{out}$ is
entangled we apply the PPT criteria for Gaussian entanglement
\cite{simon00}. For instance, we apply the positive map
$\mathcal{L}_{B}$ to the state ${\rho}_{out}$,
$\mathcal{L}_{B}({\rho}_{out})$ being the transposition (complex
conjugation) only of the subspace $B$.

Gaussian states are completely characterized by their covariance
matrix. In particular the covariance matrix of $\rho_{out}$ is
$\mathbf{V}$, with $\mathrm{V}_{\alpha\beta}=2^{-1} \langle \{
\Delta w_{\alpha},\Delta w_{\beta} \}\rangle$, where the vector
operator $\mathbf{w}=(X_{A},Y_{A}, X_{B},Y_{B})^{T}$, with the
``position''(-like) operators $X$ and ``momentum''(-like) operators
$Y$ are $X_{j}=\frac{a_{j}+ a_{j}^{\dag}}{\sqrt{2}}$
$Y_{j}=\frac{a_{j}-a_{j}^{\dag}}{i\sqrt{2}}$. Thus, the separability
properties of $\rho_{out}$ can be obtained from the positivity of
$\mathcal{L}_{B}({\rho}_{out})$, which can be expressed in terms of
its covariance matrix $\widetilde{\mathbf{V}}$ as
\begin{equation} \label{uncprinc}
\widetilde{\mathbf{V}}+\frac{i}{2}\mathbf{\Omega} \geq 0,
\end{equation}
with
\begin{equation} \label{symplectic}
\mathbf{\Omega} = \left(
\begin{array}{cccc}
0 & 1 & 0 & 0 \\
-1 & 0 & 0 & 0  \\
0 & 0 & 0 & 1  \\
0 & 0 & -1 & 0
\end{array} \right).
\end{equation}

Simon showed that $\widetilde{\mathbf{V}}$ can be calculated exactly
as $\mathbf{V}$ with a sign change in the $B$ momentum variable
(${Y}_{B}\rightarrow -{Y}_{B} $), while the other momentum and
position variables remain unchanged \cite{simon00}. Thus, we obtain
\begin{equation} \label{Vcal}
\widetilde{\mathbf{V}}= \left(
\begin{array}{cccc}
\mathcal{A}_{1} &  \mathcal{D} & \mathcal{C}_{1} &  \mathcal{G}_{1} \\
\mathcal{D}  & \mathcal{A}_{2} &  \mathcal{G}_{2}  & \mathcal{C}_{2}  \\
\mathcal{C}_{1} &  \mathcal{G}_{2}   & \mathcal{B}_{1} & \mathcal{F}   \\
 \mathcal{G}_{1}  & \mathcal{C}_{2}  & \mathcal{F} & \mathcal{B}_{2}
\end{array} \right)
\end{equation}
where in the thermal case
\begin{eqnarray}
\mathcal{A}_{1} &=&\mathcal{A}_{2}= 1/2 +\mu_A+ N (1+\mu_A+\mu_B),
\nonumber \\
\mathcal{B}_{1} &=&\mathcal{B}_{2} =1/2 +\mu_B+ N
(1+\mu_A+\mu_B), \nonumber \\
 \mathcal{C}_{1}&=&  \mathcal{C}_{2}= \sqrt{ N (N+1)}
 (1+\mu_A+\mu_B),  \nonumber \\
 \mathcal{G}_{1}&=&  \mathcal{G}_{2}= \mathcal{D}= \mathcal{F}=0.
\label{primati}
\end{eqnarray}
In the case of laser seeds the results are the same as in the case
of spontaneous PDC, i.e. the same of Eq. (\ref{primati}) with $\mu_A
= \mu_B = 0$. In the case of squeezed seeds we have
\begin{eqnarray}
\mathcal{A}_{1} &=& 1/2 +N(2+ N_A + N_B)+ \sqrt{ N_A (1+ N_A) }
(1+N)+ \sqrt{ N_B (1+ N_B) } N \cos(2\Delta \varphi) ,
\nonumber \\
\mathcal{A}_{2} &=& 1/2 +N(2+ N_A + N_B)- \sqrt{ N_A (1+ N_A) }
(1+N)- \sqrt{ N_B (1+ N_B) } N \cos(2\Delta \varphi) ,
\nonumber \\
\mathcal{B}_{1} &=& 1/2 +N(2+ N_A + N_B)+ \sqrt{ N_B (1+ N_B) }
(1+N)+ \sqrt{ N_A (1+ N_A) } N \cos(2\Delta \varphi) ,
\nonumber \\
\mathcal{B}_{2} &=& 1/2 +N(2+ N_A + N_B)- \sqrt{ N_B (1+ N_B) }
(1+N)- \sqrt{ N_A (1+ N_A) } N \cos(2\Delta \varphi) ,
\nonumber \\
\mathcal{D} &=&  \sqrt{ N_B (1+ N_B) } N \sin(2\Delta \varphi) ,
\nonumber \\
\mathcal{F} &=&  \sqrt{ N_A (1+ N_A) } N \sin(2\Delta \varphi) ,
\nonumber \\
\mathcal{C}_{1} &=& \left[1 + N_A + N_B+ \sqrt{ N_A (1+ N_A) }+
\sqrt{ N_B (1+ N_B) } \right] \sqrt{ N (1+ N) } \cos(\Delta \varphi)
,
\nonumber \\
\mathcal{C}_{2} &=& \left[-1 - N_A - N_B+ \sqrt{ N_A (1+ N_A) }+
\sqrt{ N_B (1+ N_B) } \right] \sqrt{ N (1+ N) } \cos(\Delta \varphi)
,
\nonumber \\
\mathcal{G}_{1} &=& \left[1 + N_A + N_B+ \sqrt{ N_A (1+ N_A) }
-\sqrt{ N_B (1+ N_B) } \right] \sqrt{ N (1+ N) } \cos(\Delta
\varphi) ,
\nonumber \\
\mathcal{G}_{1} &=& \left[1 + N_A + N_B- \sqrt{ N_A (1+ N_A) }+
\sqrt{ N_B (1+ N_B) } \right] \sqrt{ N (1+ N) } \cos(\Delta
\varphi),  \label{primatiSQ}
\end{eqnarray}
with $\Delta \varphi= \zeta_{A}/2+\zeta_{B}/2-\varphi $.

The PPT criterion of Eq. (\ref{uncprinc}) can be rewritten in terms
of the smallest partially transposed symplectic eigenvalue
$\widetilde{d}_{-}$ as $\widetilde{d}\geq 2^{-1}$. This condition is
never satisfied in the case of coherent and vacuum squeezed seeds,
while in the case of thermal seeds we obtain that
\begin{equation}\label{dMENO}
\widetilde{d}_{-}=\frac{1}{\sqrt{2}}\sqrt{
\mathcal{A}_{1}^2+\mathcal{B}_{1}^2+ 2 \mathcal{C}_{1}^2-
\sqrt{(\mathcal{A}_{1}+\mathcal{B}_{1})^2
[(\mathcal{A}_{1}-\mathcal{B}_{1})^2+4 \mathcal{C}_{1}^2 ]}},
\end{equation}
thus the condition $\widetilde{d}\geq 2^{-1}$ is satisfied when
\cite{DBA07}
\begin{equation} \label{autovalore}
\mu_{A} \mu_{B}- N (1+\mu_{A} + \mu_{B})\geq 0.
\end{equation}

It is noteworthy to observe that the separability/entanglement
properties of the state $\rho_{out}$ with thermal seeds can be
highlighted by the direct photon counting on A- and B- arms. In
fact, with an ideal detection system, the inequality
\begin{equation} \label{PPTn}
 \langle [\Delta
(n_{A}-n_{B})]^{2}\rangle - (\langle {n}_{A} \rangle - \langle
{n}_{B} \rangle)^2 \leq \langle {n}_{A} \rangle + \langle {n}_{B}
\rangle.
\end{equation}
exactly corresponds to Eq. (\ref{autovalore}).

Thus, as in the two previous cases we can define a parameter
quantifying the amount of the violation of the separability bound
$\mathcal{P}_{Ent}$, only in the case of thermal seeds
\begin{equation} \label{Pse1}
\mathcal{P}_{Ent}= 1- \frac{\langle [\Delta
(n_{A}-n_{B})]^{2}\rangle- (\langle {n}_{A} \rangle - \langle
{n}_{B} \rangle)^2}{\langle {n}_{A} \rangle + \langle {n}_{B}
\rangle}.
\end{equation}
$\mathcal{P}_{Ent}= 0$ corresponds to the boundary between the
separability and the entanglement regions, in fact for the state
$\rho_{out}$ with thermal seeds we obtain
\begin{equation} \label{Pse2}
\mathcal{P}_{Ent}=2 ~ \frac{ N (1+ \mu_A + \mu_B)- \mu_A \mu_B}{2 ~
N(1+ \mu_A + \mu_B)+ \mu_A + \mu_B}.
\end{equation}
According to Eq. (\ref{Pse2}) we observe that $\rho_{out}$ is
entangled when $0 < \mathcal{P}_{Ent} \leq 1$, and that the maximal
violation of the separability bound (corresponding to
$\mathcal{P}_{Ent}=1$) is achieved by the spontaneous PDC
($\mu_A=\mu_B=0$), while, if one of the two arms is seeded by the
vacuum, irrespective of the magnitude of the thermal seed on the
other arm, the state is always entangled.

We underline that the the parameter $\mathcal{P}_{Ent}$ cannot be
considered an entanglement measure (as it does not have the correct
properties) \cite{adesso}. A quantification of entanglement which
can be computed for general two-mode Gaussian states is provided by,
e.g., the logarithmic negativity.

\section{Discussion and conclusion}

\begin{figure}[tbp]
\par
\begin{center}
\includegraphics[angle=0, width=10 cm]{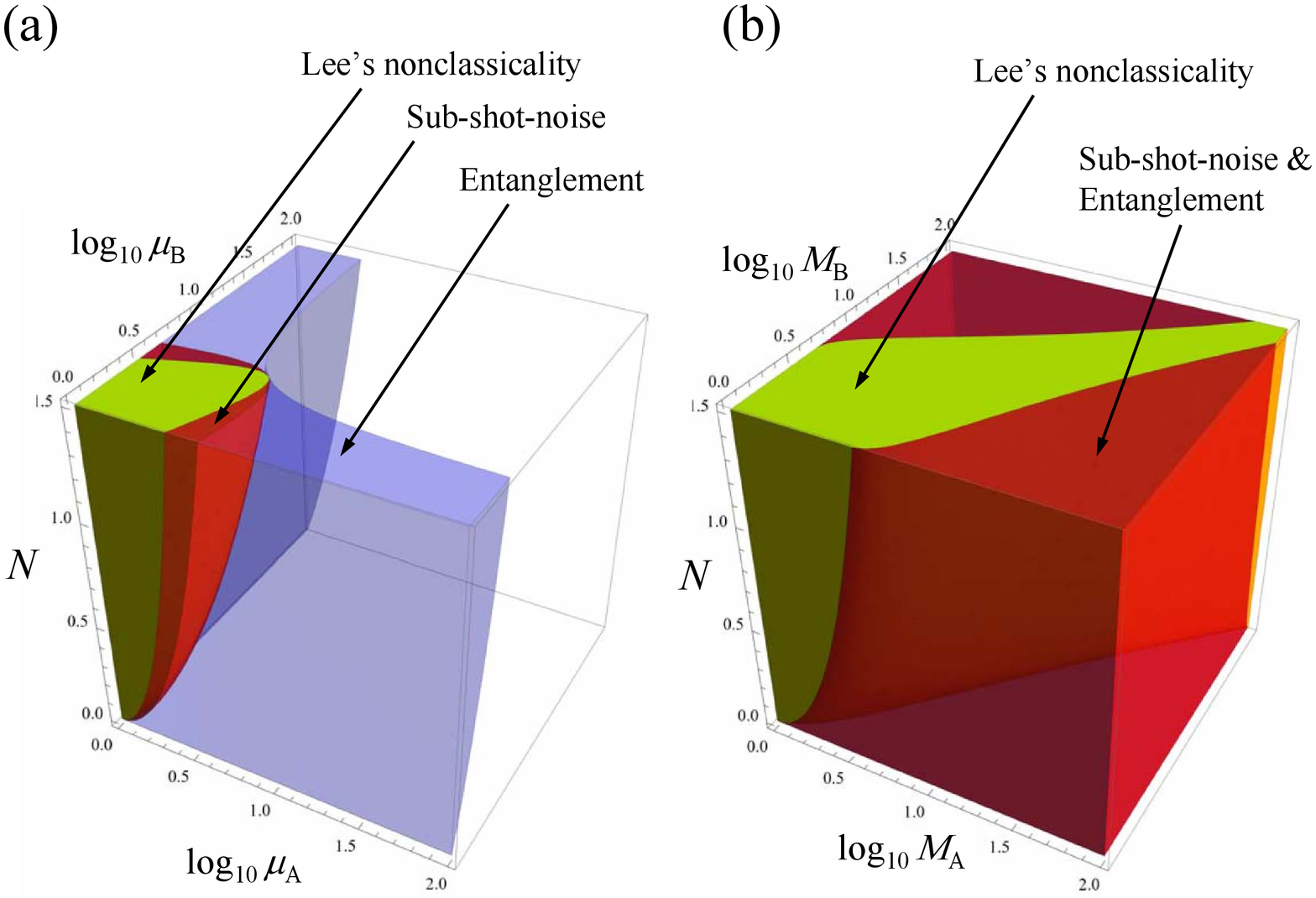}
\includegraphics[angle=0, width=10 cm]{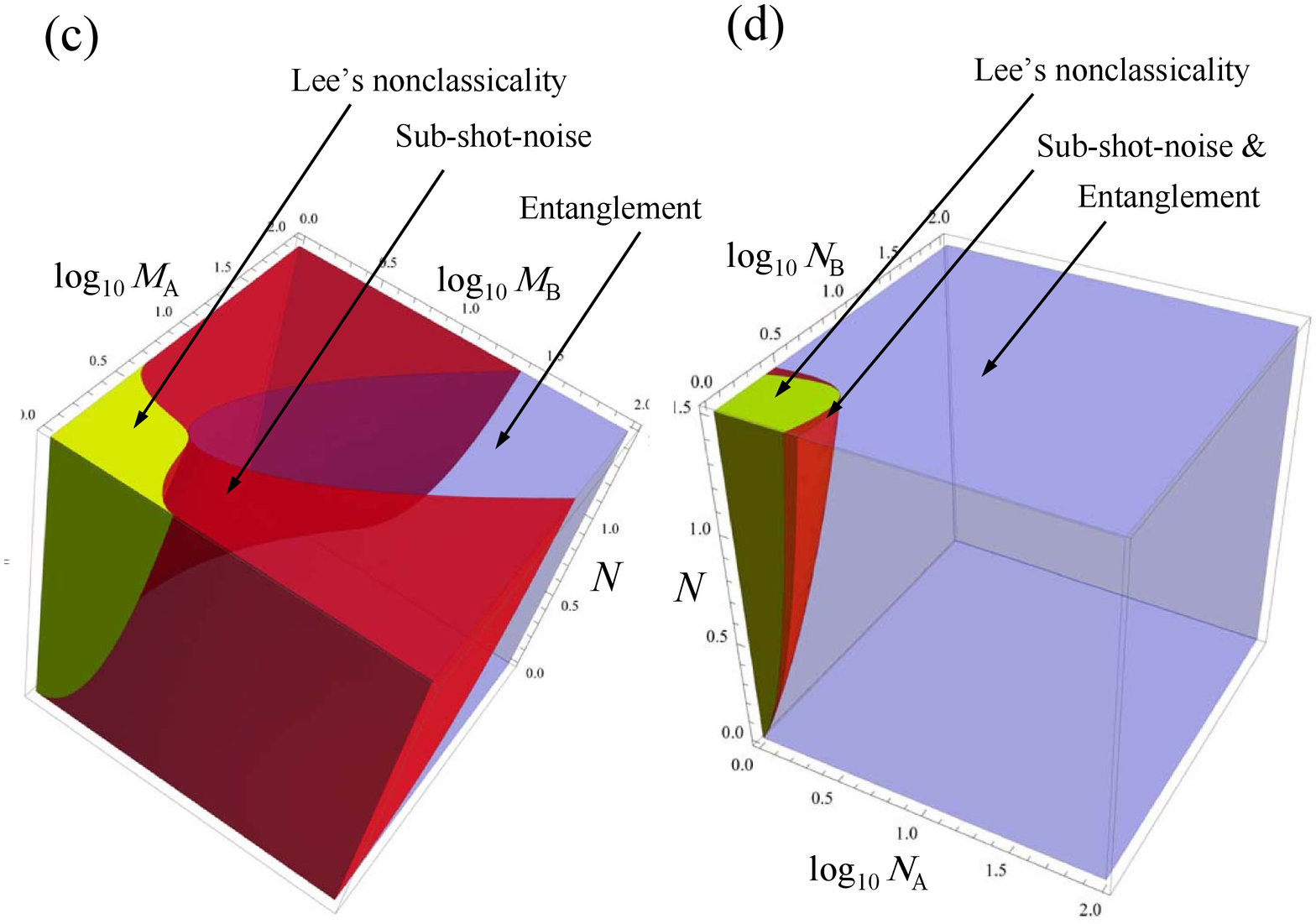}
\end{center}
\caption{ Regions of nonclassicality (entanglement, SNL violation,
Lee's classicality criterion violation) plotted as a function of the
mean number of photons of the seeding fields, and of the spontaneous
PDC ($N$) for the different state of the seeding fields considered.
In particular (a) thermal seeds, (b) coherent seeds with
$\cos(\gamma_{A}+\gamma_{B}-\varphi) = 1$, (c) coherent seeds with
$\cos(\gamma_{A}+\gamma_{B}-\varphi) = -1$, (d) vacuum squeezed
seeds.} \label{Figure 1}
\end{figure}

After having introduced the classicality quantifiers in the previous
sections, here we compare them directly: fig. 1 shows the regions of
nonclassicality of the three features considered, namely, the
entanglement, the sub-shot noise, and the Lee's nonclassicality. In
particular, in fig. 1 (a) the three regions are plotted in the case
of thermal seeds, in fig. 1 (b) and (c) are in the case of coherent
seeds with $\cos(\gamma_{A}+\gamma_{B}-\varphi)=1$ and
$\cos(\gamma_{A}+\gamma_{B}-\varphi)=-1$ respectively, and fig. 1
(d) in the case of squeezed vacuum seeds.

The most internal region corresponds to the ranges of values of seed
mean number of photons ($\mu_{A}$, $\mu_{B}$ for the thermal seeds,
$M_{A}$, $M_{B}$ for the coherent seeds, $N_{A}$, $N_{B}$ for the
squeezed vacuum seeds), and spontaneous PDC mean number of photons
$N$ where $\rho_{out}$ is nonclassical in the context of Lee's
criteria. It is noteworthy to observe that according to Eq.
(\ref{ssnA}) and Eq. (\ref{leeA}), being Lee-nonclassical is a
sufficient condition to be sub-shot noise, as it can be seen also
from Fig. 1. Furthermore, we note that in the the case of coherent
and vacuum squeezed seeds the output state is always entangled. In
the thermal case, according to Eq. (\ref{ssnA}) and to Eq.
(\ref{autovalore}), it can be observed that being sub-shot noise
limited is a sufficient condition for being entangled (fig. 1 (a)).
It is interesting to note that in the thermal case the three
conditions coincide when $\mu_{A}=\mu_{B}$. Analogously, for the
vacuum squeezed fields the SNL limit and the Lee's criterion limit
converge when $N_{A}=N_{B}$, as in can be observed in fig. 1 (d).
The same happens in the case of coherent seeds. In this case it
is interesting to note that, as the output state always violate the
SNL limit when $\cos(\gamma_{A}+\gamma_{B}-\varphi) \geq 0$, when
$M_{A}=M_{B}$ also the Lee's criterion is always satisfied (see fig.
1 (b)).

In conclusion, we have shown that there is a well defined hierarchy
among the considered nonclassicality criteria (entanglement, SNL
violation, Lee's classicality criterion violation) when PDC is seeded
by simple single mode Gaussian states (such as thermal,
coherent and vacuum squeezed states). The natural
extension of this work is the investigation of such a hierarchy in
the presence of the most general single mode Gaussian state as
seeding fields. Work along these lines is in progress and results
will be presented elsewhere \cite{degio}.




\end{document}